\documentclass[final,5p,times,twocolumn,authoryear]{elsarticle}

\usepackage{amssymb}
\usepackage{amsmath, bm, algorithm, algorithmic}
\usepackage{booktabs}

\usepackage{threeparttable}
\usepackage{multirow}

\usepackage{lineno, hyperref}

\journal{Neural Networks}
\usepackage{soul, color, xcolor}

\begin{document}

\begin{frontmatter}



\title{NSPDI-SNN: An efficient lightweight SNN based on nonlinear synaptic pruning and dendritic integration} 


\author{Wuque Cai \fnref{label1}}
\author{Hongze Sun \fnref{label1}}
\author{Jiayi He \fnref{label1}}
\author{Qianqian Liao \fnref{label1}}
\author{Yunliang Zang \fnref{label2}}
\author{Duo Chen\fnref{label1,label3}}
\author{Dezhong Yao \corref{cor1} \fnref{label1,label4}}
\author{Daqing Guo \corref{cor2}\fnref{label1,label4}}

\cortext[cor1]{dyao@uests.edu.cn}
\cortext[cor2]{dqguo@uests.edu.cn}

\affiliation[label1]{organization={Clinical Hospital of Chengdu Brain Science Institute, MOE Key Lab for NeuroInformation, China-Cuba Belt and Road Joint Laboratory on Neurotechnology and Brain-Apparatus Communication, School of Life Science and Technology, University of Electronic Science and Technology of China}, 
            city={Chengdu},
            postcode={611731}, 
            state={Sichuan},
            country={China}}
\affiliation[label2]{organization={Academy of Medical Engineering and Translational Medicine, Tianjin University}, 
            city={Tianjin},
            postcode={300072}, 
            country={China}}
\affiliation[label3]{organization={School of Artificial Intelligence, Chongqing University of Education}, 
            city={Chongqing},
            postcode={400065}, 
            country={China}}
            
\affiliation[label4]{organization={Research Unit of NeuroInformation (2019RU035)}, 
            city={Chengdu},
            postcode={611731}, 
            state={Sichuan},
            country={China}}
            

\begin{abstract}
Spiking neural networks~(SNNs) are artificial neural networks based on simulated biological neurons and have attracted much attention in recent artificial intelligence technology studies. The dendrites in biological neurons have efficient information processing ability and computational power; however, the neurons of SNNs rarely match the complex structure of the dendrites. Inspired by the nonlinear structure and highly sparse properties of neuronal dendrites, in this study, we propose an efficient, lightweight SNN method with nonlinear pruning and dendritic integration~(NSPDI-SNN). In this method, we introduce nonlinear dendritic integration~(NDI) to improve the representation of the spatiotemporal information of neurons. We implement heterogeneous state transition ratios of dendritic spines and construct a new and flexible nonlinear synaptic pruning~(NSP) method to achieve the high sparsity of SNN. We conducted systematic experiments on three benchmark datasets (DVS128 Gesture, CIFAR10-DVS, and CIFAR10) and extended the evaluation to two complex tasks (speech recognition and reinforcement learning-based maze navigation task). Across all tasks, NSPDI-SNN consistently achieved high sparsity with minimal performance degradation. In particular, our method achieved the best experimental results on all three event stream datasets. Further analysis showed that NSPDI significantly improved the efficiency of synaptic information transfer as sparsity increased. In conclusion, our results indicate that the complex structure and nonlinear computation of neuronal dendrites provide a promising approach for developing efficient SNN methods.
\end{abstract}




\end{frontmatter}



\section{Introduction}
Spiking neural networks~(SNNs) are the next generation model of artificial intelligence~(AI) and combine computational neuroscience and AI technology~\citep{maass1997networks, tavanaei2019deep}. Because the neurons of SNNs are derived from the properties of biological neurons, they have excellent energy efficiency and rich dynamic properties. These advantages make SNNs demonstrate outstanding potential in handling tasks with spatiotemporal dynamic properties. To fully exploit the potential of SNNs, researchers have developed efficient learning methods based on gradient descent in artificial neural networks~(ANNs), such as the ANN-to-SNN~\citep{wu2021tandem} and spatial-temporal backpropagation (STBP)~\citep{wu2018spatio, wu2019direct} techniques. Simultaneously, along with the development of traditional AI technology, the architectures of SNNs have also evolved from simple fully connected networks to more complex models, for example, convolutional neural networks (CNNs)~\citep{sun2023synapse}, residual networks (ResNets)~\citep{hu2021spiking}, and vision transformers~\citep{wang2023masked}. The fusion of these aspects has not only improved the processing power of SNNs, but also created new opportunities for their application in more fields, including recognition~\citep{yu2015spiking, cheng2020lisnn}, detection~\citep{kim2020spiking}, tracking~\citep{pei2019towards}, and segmentation~\citep{meftah2010segmentation}. Particularly in edge devices, SNNs can perform classical AI tasks efficiently~\citep{rathi2023exploring}.

However, simply copying the development model of ANN does not take full advantage of SNNs. Using the large model as ANNs also makes SNN encounter the problem of the computing power bottleneck. Therefore, further research needs to be considered from more dimensions. Biological neural networks are more similar to SNNs than ANNs and offer superior reliability and efficiency. Simulating the properties of biological neurons has the potential to enhance the performance of SNNs. A considerable number of brain-inspired studies have been published. Among these, a particularly effective method inspired by neuronal heterogeneity involves parameterizing or dynamizing the membrane potential constants~\citep{fang2021incorporating} or firing thresholds~\citep{sun2023synapse, ding2022biologically} of leaky integrate-and-fire~(LIF) neurons~\citep{dayan2003theoretical}. Researchers have shown that, unlike the dendrites of simple point neurons, those of pyramidal neurons enable more complex functions, such as "XOR" arithmetic~\citep{gidon2020dendritic}. In additional studies, researchers have shown that the activity of individual pyramidal neurons is comparable to that of deep neural networks~\citep{beniaguev2021single}. Therefore, a practical approach to improve deep neural networks may be to model the complex structure and functions of dendrites.

\begin{figure*}[t]
	\centering
	\includegraphics{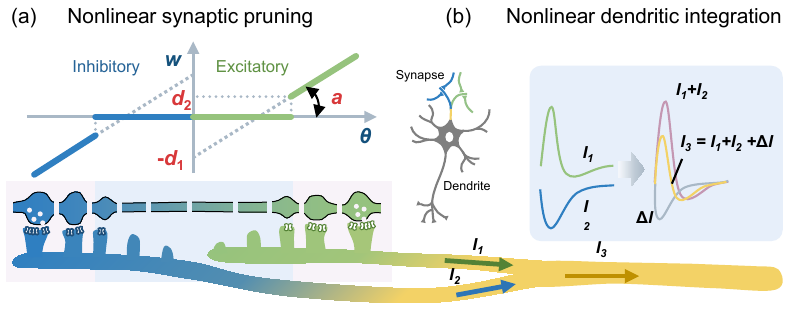}
	\caption{Schematic illustration of nonlinear state transitions in dendritic spines and dendritic integration in biological neurons.
		(a) State transitions between mature dendritic spines and filopodia, as well as between excitatory and inhibitory synapses in biological neurons. Inhibitory synapses (left, blue) utilize $\gamma$-aminobutyric acid (GABA) as their primary neurotransmitter, whereas excitatory synapses (right, green) rely on glutamate. The nonlinear weight reparameterization, inspired by these synaptic transitions, is depicted in the function plot in the upper left.
		(b) Dendritic integration of synaptic inputs is inherently nonlinear. The resulting membrane potential is modeled as the linear summation of input potentials augmented by a nonlinear correction term.}
	\label{Fig1}
\end{figure*}

In recent studies, researchers have found that the nonlinear computation of dendrites plays a crucial role in modeling biological neurons. This capability to integrate nonlinear information ensures that biological neural networks perform effectively, despite being highly sparse~\citep{hao2009arithmetic, li2014bilinearity, li2019dendritic}. In experiments and modeling studies involving rat hippocampal medullary slices, researchers discovered the presence of nonlinear product components that integrate excitatory glutamate inputs and inhibitory GABA inputs. The researchers measured these inputs and derived approximate mathematical models~\citep{hao2009arithmetic}. A bilinear spatiotemporal dendritic integration rule for all types of multiple synaptic inputs was derived in a more in-depth asymptotic analysis of the two-compartment passive cable neuron model, as shown in Fig.~\ref{Fig1}(b). The nonlinear coefficients were associated with both the input time and input location. The researchers validated the findings of this study by simulating a real vertebral neuron model and conducting electrophysiological experiments on rat hippocampal CA1 neurons~\citep{li2014bilinearity}. In recent experiments, researchers reliably parameterized the role of each pair of significant inputs.
By deriving a bilinear dendritic integration rule for multiple synapses, the bilinear rule provided higher computational power for point neurons. The experiments also demonstrated the highly spatiotemporal dependence of bilinear coefficients within the input information~\citep{li2019dendritic}. Therefore, incorporating the nonlinear dendritic integration~(NDI) of the bilinear rule into the neurons of the SNN can improve the spatiotemporal properties of the SNN. 

The increase in SNN nonlinearity is accompanied by an increase in learnable parameters and the hardware burden. Because of the originally required huge computing power support of SNNs, it is imperative to make the SNN models lightweight. The development of lightweight models has proven to be very effective in deep learning because it solves the challenge of running high-performance models under limited hardware resources~\citep{wang2022lightweight}. These lightweight methods include pruning~\citep{liang2021cemodule, chen2020dynamical}, tensor decomposition~\citep{phan2020tensor}, quantization~\citep{fei2021general}, and knowledge distillation~\citep{ding2022dual}. In most of these methods, mathematical and engineering aspects have been emphasized primarily. They lack biologically rational foundations and exhibit drawbacks, such as suboptimal optimization and limited generalizability. By contrast, pruning involves reducing connections, which closely mimics the evolution of biological neural networks. Additionally, unstructured pruning methods are well-suited to SNNs because hardware computations occur only when transmission spikes occur and the connection weights are non-zero~\citep{chen2022state}. Consequently, it is necessary to develop a biologically based unstructured pruning method.

A recent biologically inspired pruning technique, known as the state transition of dendritic spines~(STDS), simulates synaptic sparsification by reparameterizing connection weights~\citep{chen2022state}. STDS is inspired by the state transitions between mature dendritic spines-filopodia and excitatory-inhibitory synapses observed in biological neurons, especially in Purkinje cells. In our approach, we further extend this concept by explicitly modeling the heterogeneity of state transition ratios (~\citep{dobrunz1997heterogeneity, fritschy2012molecular, perez2021neural}) in different neurons and neuronal populations. This neurobiological variability is a foundation for developing a more flexible and adaptive pruning mechanism, as illustrated in Fig.~\ref{Fig1}(a). To further enhance sparsity, we incorporate a threshold-based pruning mechanism with the state transition ratio. This dual-pruning strategy enables the synaptic reparameterization process to adapt more flexibly to structural heterogeneity, allowing the network to learn diverse and effective patterns of synaptic transformation grounded in biological plausibility.

Accordingly, this study introduces a nonlinear synaptic pruning and dendritic integration method for constructing efficient, lightweight SNNs based on nonlinear synaptic pruning and dendritic integration~(NSPDI-SNN). The proposed approach enhances dendritic computational power through the integration of the nonlinear dendritic integration~(NDI) method.  Additionally, we present an NSP method that simulates the heterogeneous transition ratios of dendritic spines and filopodia. This method uses a flexible threshold-setting strategy by aiding adjustments to the sparsification goals. By combining the above methods, a tunable sparsification framework is developed, enabling improved performance at high sparsity levels. The main contributions of this study are summarized as follows:
\begin{itemize}
	\item [$\bullet$] We introduce NDI into SNNs to enhance the ability of models to represent spatiotemporal information.
	
	\item [$\bullet$] By considering the heterogeneity and fusing NDI, we propose an NSPDI method that can effectively sparsify SNNs and maintain high performance.
	
	\item [$\bullet$] We illustrate that the transition gain, transition thresholds, and pruning thresholds in the NSPDI method improve the flexibility and controllability of the weight distribution to achieve sparsification.
	
	\item [$\bullet$] In extensive experiments, we demonstrated that the proposed model achieved high sparsity with low accuracy losses across four datasets.
\end{itemize}

The remainder of this paper is organized as follows. In Section II, we summarize the nonlinear integration methods and unstructured sparsification methods. In Section III, we provide a comprehensive analysis of the NSPDI-SNN model and the associated learning techniques. Section IV presents a comprehensive experimental analysis of the proposed method using various datasets. Finally, in Section V, we provide a concise discussion and conclusion.

\section{Related Work}
\subsection{Nonlinear Dendritic Integration}
Dendrites provide neurons with powerful computational capabilities. Learning from dendritic structures is an approach to improve AI performance. One method is polynomial neural networks~\citep{ma2005constructive, jiang2016potential}, which integrates high-order function formulas into neuron activation functions, thereby improving the nonlinearity of neurons. Another method directly introduces higher-order term relationships between input information to replace the traditional neuron structure~\citep{chen2022polynomial, liu2021dendrite, NEURIPS2024_90b31ad3}. However, this approach has so far only been validated on simple datasets and requires further investigation on more complex datasets. A common approach is incorporating nonlinear calculations into the input integration process~\citep{mantini2021cqnn, xu2022quadralib, zheng2024temporal}. Nonlinear calculations greatly increase the computational complexity and parameter amount of the model and can improve performance on traditional datasets. In recent experimental calculations and observations, the researchers have found that dendritic computations that involve spiking adhere to a bilinear process~\citep{hao2009arithmetic, li2014bilinearity, li2019dendritic}. Based on this discovery, we propose integrating the bilinear property into SNN computations to enhance computational performance when a highly sparse model is preserved.

\subsection{Synaptic Pruning}
Pruning is a crucial aspect of model compression. The process of sparsifying weights is similar to biological neuronal synaptic pruning, which aligns with the information transfer procedure of SNN models. Following the successful application of traditional ANN pruning, SNN pruning is typically integrated with learning. The spike-timing-dependent plasticity learning method relies on activity correlations among neurons, which enables the straightforward determination of correlations between the connected neurons. It can implement the synapse pruning process based on this information~\citep{rathi2018stdp, nguyen2021connection}. The ANN-to-SNN conversion method initially trains an ANN model using the same structure and subsequently eliminates the redundant synapses based on predefined rules derived from the weights and distributions of the pre-trained model~\citep{mern2017layer, liu2019application}. More successful STBP training uses a substitution function to solve the non-conductivity problem during the spike firing process, which enables the developed SNN to be pruned in a gradient-based framework~\citep{deng2021comprehensive, yin2021energy}. Moreover, a range of pruning techniques inspired by biological neuron regeneration has emerged within the learning framework of gradient descent~\citep{chen2021pruning}. These methods aim to balance sparsity and accuracy. Another STDS method that simulates dendritic spine transition~\citep{chen2022state, deng2021comprehensive} also successfully achieves high-sparsity pruning for ResNet and is the basis of our proposed model.

\section{Method}
\subsection{Nonlinear Integration in Dendrites}
The LIF model is the most common neuron model in SNNs and provides both algorithmic convenience and rich dynamics. In SNNs, the iterative formula of a neuron can usually be expressed as follows:
\begin{equation}
	\begin{aligned}
		u_t &= \left(1-\frac{1}{\tau_m} \right)u_{t-1}\left(1-s_{t-1}\right) + I_t,\\
		s_t &= g\left(u_t\right)=
		\begin{cases}
			1, \quad u_t\ge u_{\mathrm{th}}\\
			0, \quad u_t < u_{\mathrm{th}}
		\end{cases},
	\end{aligned}
	\label{Eq1}
\end{equation}
where $u_t$, $s_t$, and $I_t$ are the membrane potential, spike, and input current, respectively, of the neuron at time $t$ and $\tau_m$ is the membrane time constant. $g(\cdot)$ is the Heaviside function, which describes the process of spike firing. When $u_t$ reaches the membrane potential $u_{\mathrm{th}}$, a spike $s_t$ of 1 is generated, and $u_t$ is reset to the resting potential of 0 mV.

However, in contrast to the linear integration of the input currents in a typical SNN neuron, the dendritic integration process in biological neurons is nonlinear. This ensures that dendrites are extremely computationally powerful. Specifically, the input current of a LIF neuron can be given as $I_t = \sum_{i}ws_i$, which is a simple dot product calculation; the input current of a biological neuron involves a highly complex integration process. Researchers have demonstrated experimentally that, for the dendritic computing model, dendritic integration conforms to bilinearity. As shown in Fig.~\ref{Fig1}(b), when two dendritic currents are pooled together, an additional nonlinear term is added to the amplitude of the obtained current. The summed dendritic potential $u_\mathrm{S}$ for two effective excitatory and inhibitory inputs can be described as follows:
\begin{equation}
	u_\mathrm{S} = u_\mathrm{E} + u_\mathrm{I} + k_\mathrm{EI} \cdot u_\mathrm{E}u_\mathrm{I}, 
	\label{Eq2}
\end{equation}
where $u_\mathrm{E}$ and $u_\mathrm{I}$ denote the potentials induced by excitatory and inhibitory inputs, respectively, and $k_\mathrm{EI}$ represents the correction coefficient.

\begin{figure}[t]
	\centering
	\includegraphics{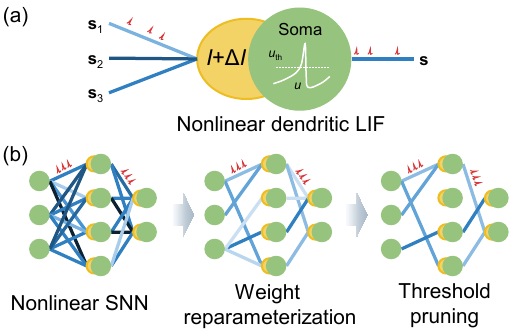}
	\caption{NDI and NSP. (a) An NDI-LIF model. When input spiking trains ($\mathbf{s}_1$, $\mathbf{s}_2$, $\mathbf{s}_3$) are fed into neurons, the current is integrated with the correction term $\Delta I$ (yellow). Then, through the soma (green), an output spiking train ($\mathbf{s}$) is issued. (b) Complete scheme of NSP. First, an initial SNN model consisting of NDI-LIF neurons. The model's weights are reparametrized to complete the initial pruning. Then, a threshold pruning strategy is used to reduce the weighted connections further.}
	\label{Fig2}
\end{figure}

Therefore, the above theory can improve the input integration process for LIF neurons, as shown in Fig.~\ref{Fig2}(a). Mathematically, for a fully connected layer with $M$ neurons, the NDI of $N$ inputs $\bm{x} \in \mathbb{R}^{N}$ can be described as follows:
\begin{equation}
	\bm{I} = \bm{W}\bm{x} + \bm{b} + \bm{K}\bm{x}^2,
	\label{Eq3}
\end{equation}
where $\bm{I} \in \mathbb{R}^{M}$ denotes the input current obtained through dendritic integration, $\bm{W} \in \mathbb{R}^{M \times N}$ is the synapse connectivity weights, $\bm{b} \in \mathbb{R}^{M}$ is the bias vector and $\bm{K} \in \mathbb{R}^{M \times N}$ are nonlinear integration weights. Here, the form of $\left(\bm{W}\bm{x}\right) \odot \left(\bm{V}\bm{x}\right)$ is used to the quadratic term:
\begin{equation}
	\bm{I} = \bm{W}\bm{x} + \bm{b} + \left(\bm{W}\bm{x}\right) \odot \left(\bm{V}\bm{x}\right),
	\label{Eq4}
\end{equation}
where $\odot$ is Hadamard product, $\bm{V}$ is another parameter that controls $\bm{K}$ together with $\bm{W}$, which can be considered from the synapse-level, neuron-level, and layer-level:
\begin{itemize}
	\item \textbf{Synapse-level}. For $\bm{V} \in \mathbb{R}^{M \times N}$, the quadratic term computes pairwise feature interactions $\left(\bm{W}\bm{x}\right) \odot \left(\bm{V}\bm{x}\right)$.
	\item \textbf{Neuron-level}. For $\bm{v} \in \mathbb{R}^{M}$, a diagonal matrix transforms $\bm{K}$ into a rank-1 interaction matrix $\left(\bm{W}\bm{x}\right) \odot \left(\bm{v} \odot \mathbf{1}^{M \times N}\bm{x}\right)$.
	\item \textbf{Layer-level}. For $v \in \mathbb{R}$,  the nonlinear term simplifies to global feature scaling $v\left [ \left(\bm{W}\bm{x}\right) \odot \left(\mathbf{1}^{M \times N}\bm{x}\right)\right ]$.
\end{itemize}
The synapse-level parameter $\bm{V}$ captures the most granular nonlinear integration dynamics, enabling the richest representation of neuronal activity (e.g., branch-specific dendritic interactions), but comes with the highest computational cost (parameter complexity O(MN)). The layer-level scalar $v$ models global nonlinear integration (e.g., neuromodulatory effects) with minimal computational overhead (parameter complexity O(1)), though it lacks fine-grained control. In contrast, the neuron-level vector $\bm{v}$ achieves an optimal balance – maintaining biologically plausible somatic-dendritic heterogeneity through per-neuron nonlinear gain modulation (parameter complexity O(M)), while keeping computational complexity tractable (same order as linear layers). This provides an efficient trade-off between representational capacity and computational efficiency.

Extended to convolutional neural networks, NDI has the same form:
\begin{equation}
	\bm{I} = \bm{W} \circledast \bm{x} + \bm{b} + \left(\bm{W} \circledast \bm{x}\right) \odot \left(\bm{V} \circledast  \bm{x}\right),
	\label{Eq5}
\end{equation}
where $\circledast$ is the convolution operation, $\bm{x}$ denotes neural inputs with $C \times H \times W$, where $C$ is the number of channels, and $H$ and $W$ are the height and width of the inputs, respectively. $\bm{W} \in \mathbb{R}^{C_\text{out} \times C_{\text{in}} \times H \times W}$ represent the convolutional weight and $\bm{b} \in \mathbb{R}^{C_\text{out} \times 1 \times 1 \times 1}$ represent the convolutional bias. Similarly, $\bm{V}$ can be divided into three levels: synapse-level ($\bm{V} \in \mathbb{R}^{C_\text{out} \times C_{\text{in}} \times H \times W}$), channel-level ($\bm{V} = \bm{v} \odot \mathbf{1}^{M \times N}$, where $\bm{v} \in \mathbb{R}^{C_\text{out} \times 1 \times 1 \times 1}$) and layer-level ($\bm{V} = v\mathbf{1}^{M \times N}$, where $v \in \mathbb{R}$).

Pruning remains essential in SNNs even after incorporating channel-level nonlinear integration, as it mitigates the computational inefficiencies and redundancy introduced by nonlinear operations while preserving the model's event-driven sparsity and energy efficiency. By strategically removing superfluous connections, pruning not only maintains the enhanced representational capacity afforded by nonlinearity but also optimizes the network structure for neuromorphic hardware implementation, demonstrating that these two techniques are fundamentally complementary in achieving both expressive power and operational efficiency.

\subsection{Nonlinear Synaptic Pruning}
\subsubsection{State Transition of Heterogeneous Dendritic Spines}
STDS provides a weight reparameterization framework for dynamically pruning SNN models. The relationship between the real connection weight $w$ and reparameterization parameters $\theta$ is characterized as follows:
\begin{equation}
	w=\operatorname{sign}(\theta) \cdot(|\theta|-d)_{+}, d \geq 0,
	\label{Eq6}
\end{equation}
where $(x)_{+}=\operatorname{max}(x,0)$, $\operatorname{sign}(x)$ is the symbolic function, and $d$ is the transition threshold that controls the sparsity of the true connection weights. Each reparametrized connection parameter below the value of parameter $d$ is treated as a filopodium. Predictably, assuming a normal distribution for the reparameterization parameters $\theta$, synapses are gradually converted to filopodia as the value of $d$ increases and the sparsity level increases.

Because of inherent heterogeneity among individual neurons and neuronal populations, the development and transformation processes of neuronal synapses exhibit some diversity, both among neurons and across synaptic connections. In this study, we introduce an transition gain to represent the state heterogeneity of dendritic spines. The transition gain indicates the efficiency of heterogeneous state transition and participate in the reparameterization of the synaptic development and transformation procedures. The reparameterization of heterogeneous state transition can be expressed as
\begin{equation}
	w_p = \operatorname{sign} \left( \theta \right) \cdot \left [a \left( |\theta|-d_1 \right) \right ]_{+},  d_1 \ge 0,
	\label{Eq7}
\end{equation}
where $w_p \in \mathbb{R}$ is the post-transition weight. $a > 0$ is the learnable transition gain. It shows how the parameter $\theta \in \mathbb{R}$ affects the real connection weight $w_p$. The parameter $d_1 \in \mathbb{R}$ signifies the transition threshold. Here, $\bm{\Theta}$ denotes the set of all learnable parameters $\theta$. The $a$ can be hierarchically defined at three levels, analogous to nonlinear integration: synapse-level, channel-level (neuron-level for fully connected layers), and layer-level. Its mathematical formulation follows a similar structure to the nonlinear integration weight $\bm{V}$, and thus, for brevity, we omit the detailed derivation here.

Connections with values exceeding threshold $d_1$ (i.e., $\{\theta \in \bm{\Theta} \mid \theta > d_1\}$)) are classified as dendritic spine connections, while the remaining connections  correspond to filopodia. Increasing $d_1$ reduces the proportion of mature spines and decreases the connection weights, thus the network becomes sparser. Conversely, when $d_1$ decreases, the proportion of dendritic spines increases and the connection weights also increase, thus the network becomes denser. 

The strength of synaptic connections varies because of differences in $a$. In our design, $a$ is strictly positive to ensure that the post-transition weight $w_p$ preserves the correct sign and magnitude relative to $\theta$. During training, we enforce this constraint by clamping the updates of $a$, preventing it from becoming negative. This natural variation lets the remaining connections (after pruning) make up for lost information by adjusting their strengths, similar to how synaptic scaling works in real neurons to keep networks stable.

\subsubsection{Threshold Pruning}
Additionally, for synaptic connections with tiny actual weights, their information transfers can be directly ignored, as shown in Fig.~\ref{Fig2}(b). In this case, a pruning threshold is added to cut off small weight connections:
\begin{equation}
	\begin{aligned}
		&w =\begin{cases}
			w_p, & |w_p| > d_2\\
			0, & |w_p| \le d_2
		\end{cases}, d_2 \ge 0,
		\label{Eq8}
	\end{aligned}
\end{equation}
where $d_2$ is the pruning threshold, below which the weights are considered negligible. Substituting Eq.~(\ref{Eq8}) into Eq.~(\ref{Eq7}) allows for the derivation of the final reparametrized function of the connection weights in the SNN model. For a weight $w \in \bm{W_{\mathrm{p}}}$ can be derived as follows:
\begin{equation}
	\begin{aligned}
		w =\begin{cases}
			\operatorname{sign} \left( \theta \right) \cdot \left [ a \left(|\theta|-d_1 \right) \right ] _{+}, & |\theta| > d_1 +  \frac{d2}{a}\\
			0, & |\theta| \le d_1 + \frac{d2}{a}
		\end{cases}.\\
	\end{aligned}
	\label{Eq9}
\end{equation}

As seen in the transition curves exhibited by $\theta$ and $w$ in Fig.~\ref{Fig1}(a), the weight transition function is a segmented function, with $a$, $d_1$, and $d_2$ collectively determining the locations of the segmented points. As a result, the transition function is more flexible and better controls the sparsification and distribution of $w$.

\subsection{Learning Method}
The STBP learning method is uesd to train the model. To address the non-differentiable Heaviside function $g(\cdot)$, we use the surrogate function for approximation purposes,
\begin{equation}
	g(u_t) = \frac{1}{\pi} \operatorname{arctan} \left[ \pi \left( u_t - u_{\mathrm{th}} \right) \right] + \frac{1}{2}.
	\label{Eq10}
\end{equation}
As a result, the spatiotemporal gradient information used for optimization can be passed backward through the LIF neurons.

In this study, the mean squared deviation is used as the loss function, which is described as follows:
\begin{equation}
		\mathcal{L} = \frac{1}{2N} \sum_{n=1}^{N} \left\| \mathbf{y}_n - \frac{1}{T} \sum_{t=1}^{T} \mathbf{s}_{t, n}^{o} \right\|_2^2.
	\label{Eq11}
\end{equation}
The loss value $\mathcal{L}$ is calculated from the $N$ training samples. For each training sample, the square error between the corresponding predicted value $\sum_{t=1}^{T} \mathbf{s}_{t, n}^{o}$ and the label $Y_N$ is calculated. The predicted values account for the classification neuron spikes $\mathbf{s}_{t, n}^{o}$ in $T$ time windows.

Along the path through the computational graph of the model, the gradient of $\mathcal{L}$ with respect to $I$ can be computed:
\begin{equation}
	\begin{aligned}
		\nabla_{I_t}\mathcal{L} &= \frac{\partial \mathcal{L}}{\partial I_t} = \frac{\partial \mathcal{L}}{\partial u_t} \cdot \frac{\partial u_t}{\partial I_t}.\\
	\end{aligned}
	\label{Eq12}
\end{equation}
Furthermore, the gradients of the parameters in the nonlinear integration ($\nabla_{w} \mathcal{L}$, $\nabla_{v} \mathcal{L}$ and $\nabla_{b} \mathcal{L}$) are:
\begin{equation}
	\begin{aligned}
		\nabla_{w} \mathcal{L} &= {I} \mathcal{L} \cdot \frac{\partial I}{\partial w} = \nabla_{I} \mathcal{L} \cdot x +  \left(v \cdot x\right) \cdot x \cdot \nabla_{I} \mathcal{L},\\
		\nabla_{v} \mathcal{L} &= \nabla_{I} \mathcal{L} \cdot \frac{\partial I}{\partial v} = \left(w \cdot x\right) \cdot x \cdot \nabla_{I} \mathcal{L},\\
		\nabla_{b} \mathcal{L} &= \nabla_{I} \mathcal{L} \cdot \frac{\partial I}{\partial b} = \nabla_{I} \mathcal{L}.\\
	\end{aligned}
	\label{Eq13}
\end{equation}

\begin{table*}[!t]
	\caption{Computational complexity comparison of different modules.
	\label{Tab1}}
	\centering
	\setlength{\tabcolsep}{5.0mm}{
		\begin{tabular}{l|cc}
			\hline
			Module       &          Computing Complexity          &    Number of parameters    \\ \hline
			Linear       &              $O(MN)$              &        $O(MN)$         \\
			NDI-Linear   &             $O(2\cdot MN + 2M)$              &     $O(2\cdot MN + M)$     \\
			NSPDI-Linear &          $O(2\beta\cdot MN + 2M)$           &   $O(2\beta\cdot MN + 3\cdot M)$   \\ \hline
			Conv         &       $O(C_{i}C_{o}K^2HW)$        &     $O(C_{i}C_{o}K^2)$      \\
			NDI-Conv     & $O(2\cdot C_{i}C_{o} K^2HW + 2\cdot C_{o}HW)$ & $O(2\cdot C_{i}C_{o}K^2 + C_{o})$ \\
			NSPDI-Conv   &$O(2\beta\cdot C_{i}C_{o} K^2HW + 2\cdot C_{o}HW)$&  $O(2\beta\cdot C_{i}C_{o}K^2 + 3\cdot C_{o})$\\ \hline
		\end{tabular}
	}
\end{table*}

Based on Eqs.~(\ref{Eq7}) and (\ref{Eq8}), gradient descent can directly optimize $\theta$ and $a$. The gradient of the reparameterization parameter $\nabla_{\theta} \mathcal{L}$ and the gradient of the transition gain $\nabla_{a} \mathcal{L}$ are zero when $ \left | \theta \right| \leq d_1 + \frac{d_2}{a}$. In all other cases, they can be described as follows:
\begin{equation}
	\begin{aligned}
		\nabla_{a_w}\mathcal{L}&=\nabla_{w}\mathcal{L}\cdot\frac{\partial w}{\partial a_w}=\nabla_{w} \mathcal{L} \cdot \theta_w ,\\	\nabla_{\theta_w}\mathcal{L}&=\nabla_{w}\mathcal{L}\cdot\frac{\partial w}{\partial \theta_w}=\nabla_{w} \mathcal{L} \cdot a_w,
	\end{aligned}
	\label{Eq14}
\end{equation}
and
\begin{equation}
	\begin{aligned}
		\nabla_{a}\mathcal{L}&=\nabla_{v}\mathcal{L}\cdot\frac{\partial v}{\partial a}=\nabla_{v} \mathcal{L} \cdot \theta_v,\\
		\nabla_{\theta}\mathcal{L}&=\nabla_{v}\mathcal{L}\cdot\frac{\partial v}{\partial \theta}=\nabla_{v} \mathcal{L} \cdot a_v.
	\end{aligned}
	\label{Eq15}
\end{equation}
where $\theta_w$ and $\theta_v$ are the linear reparameterization weight and nonlinear reparameterization weight, respectively, and $a_w$ and $a_v$ are the linear transition gain and nonlinear transition gain, respectively.
\begin{algorithm}[!t]
	\caption{Training procedure of NSPDI-SNN model}
	\label{alg:nspdi_snn}
	\textbf{Input}: Training dataset $\mathcal{D}_{train}$, network structure specification $\mathcal{S}$, transition threshold $d_{1}$, pruning threshold $d_{2}$\\
	\textbf{Parameter}: Linear reparameterization weight $\theta_w$, nonlinear reparameterization weight $\theta_v$, transition gain $a_w$ and $a_v$, batch normalization $\mathcal{BN}$ and loss function $\text{MSELoss}$\\
	\textbf{Output}: Trained NSPDI-SNN model
	\begin{algorithmic}
		\STATE Initialize model $\mathcal{M}$ with NSPDI-SNN layers from $\mathcal{S}$
		\STATE Load $\mathcal{D}_{train}$
		\FOR{epoch}
		\STATE Update $d_{1},d_{2}$ according to epoch
		\FOR{mini-batch $(x,y) \in \mathcal{D}_{train}$}
		\FOR{each layer $\ell$ in $\mathcal{M}$}
		\STATE $w \gets \text{NSP}(\theta_w, a_w, d_{1}, d_{2})$, \;
		$v \gets \text{NSP}(\theta_v, a_v, d_{1}, d_{2})$ \COMMENT{NSP reparameterization}. Eq.~(\ref{Eq9})
		\STATE $I \gets \text{NDI}(x, w, v, b)$ \COMMENT{NDI integration}. Eq.~(\ref{Eq4})
		\STATE $I \gets \mathcal{BN}(I)$ \COMMENT{Batch normal}.
		\STATE $(s,u) \gets \text{LIF}(I, s, u)$ \COMMENT{LIF neuron update}. Eq.~(\ref{Eq1})
		\ENDFOR
		\STATE $I^o \gets \text{Linear}(s, w, b)$
		\STATE $(s^{o},u^{o}) \gets \text{LIF}(I^o, s, u^{o})$ \COMMENT{Spiking outputs}. Eq.~(\ref{Eq1})
		\STATE $\mathcal{L} \gets \text{MSELoss}(s^{o}, y)$. Eq.~(\ref{Eq11})
		\STATE Backpropagate $\nabla \mathcal{L}$ and update. Eq.~\ref{Eq10} and Eqs.~(\ref{Eq12})-(\ref{Eq15})
		\ENDFOR
		\ENDFOR
	\end{algorithmic}
	\label{arg1}
\end{algorithm}

Using Eqs.~(\ref{Eq10})-(\ref{Eq15}), NSPDI-SNN enables efficient pruning during training. The overall training procedure is summarized in Algorithm~\ref{arg1}. When the NSP or NDI method is removed, our model degenerates into a dense SNN based on NDI~(NDI-SNN) or lightweight SNN based on NSP~(NSP-SNN), respectively.


\begin{table*}[!t]
	\renewcommand{\arraystretch}{1.25}
	\caption{Default values of the hyperparameters for different datasets.}
	\label{Tab2}
	\centering
	\setlength{\tabcolsep}{2.0mm}{
		\begin{tabular}{l|ccccccccccc}
			\hline
			Dataset        & Trial & Epoch & BS  & $\eta$ & $\gamma_{\eta}$ & $u_{\mathrm{t}h}$ & $\tau_m$ & $T$ & $\Delta t(\mathrm{ms})$ & $\text{Epoch}_{\text{C}}$ & $T_{\text{cyc}}$ \\ \hline
			DVS128 Gesture &  10   &  500  & 32  &  1e-3  &     Cosine      &        1.0        &   3.3    & 10  &           125           &            50             &        50        \\
			CIFAR10-DVS    &  10   &  200  & 80  &  2e-4  &     Cosine      &        1.0        &   2.0    & 20  &           50            &            25             &        25        \\
			CIFAR10        &   3   &  400  & 100 &  5e-2  &     Cosine      &        1.0        &   2.0    &  4  &            -            &            50             &        25        \\
			SHD            &  10   &  200  & 128 &  1e-2  &     Cosine      &        0.3        &   2.0    & 100 &           14            &            25             &        25        \\
			Maze2D         &  10   &   -   & 100 &  1e-3  &        -        &        0.5        &   2.0    &  4  &            -            &            25             &        25        \\ \hline
		\end{tabular}
	}
\end{table*}

\begin{table*}[!t]
	\renewcommand{\arraystretch}{1.25}
	\caption{Architectures of SNN models.}
	\label{Tab3}
	\newcommand{\tabincell}[2]{
		\begin{tabular}
			{@{}#1@{}}#2
		\end{tabular}
	}
	\centering
	\setlength{\tabcolsep}{4mm}{
		\begin{tabular}{l|l}
			\hline
			Dataset        & Network architecture                                                                            \\ \hline
			DVS128 Gesture & \tabincell{l}{Inputs-128C5S2-BN-AP2-128C3-BN-AP2-128C3-BN-AP2-128C3-BN-AP2\\-128C3-BN-MP2-FC11} \\ \hline
			CIFAR10-DVS    & \tabincell{l}{Inputs-128C7S2-AP2-128C3-AP2-128C3-AP2-128C3-AP2-128C3-MP2-FC10}                  \\ \hline
			CIFAR10        & ResNet19                                                                                        \\ \hline
			SHD            & Inputs-FC128-DP-FC128-DP-FC20                                                                   \\ \hline
			Maze2D         & Inputs-FC256-FC256-FC4                                                                          \\ \hline
		\end{tabular}
	}
\end{table*}

\subsection{Computational Complexity Analysis}
To systematically evaluate the computational efficiency of the proposed NSPDI method, we conduct a theoretical analysis of its computational complexity and parameter count. As shown in Tab.~\ref{Tab1}, the analysis reveals the inherent characteristics and potential advantages of the NSPDI module in terms of computational resource consumption.

The results indicate that the NSPDI module effectively reduces the computational burden for sparsity $\beta \in (0,1]$. Specifically, for the convolution module, the computational complexity of NSPDI-Conv is $O(2\beta \cdot C_i C_o K^2 HW + 2 \cdot C_o HW)$, which reduces the dominant term of the dual-path convolution by approximately $1/\beta$ compared with NDI-Conv's $O(2 \cdot C_i C_o K^2 HW + 2 \cdot C_o HW)$, since each path only processes $\beta \cdot C_i$ input channels. When $\beta < 0.5$, the theoretical computational cost of NSPDI-Conv is significantly lower than that of NDI-Conv. Regarding the number of parameters, NSPDI-Conv requires $O(2\beta \cdot C_i C_o K^2 + 3 \cdot C_o)$, also approximately $1/\beta$ times smaller than NDI-Conv's $O(2 \cdot C_i C_o K^2 + C_o)$. NSPDI-Linear exhibits a similar optimization trend for linear modules. In summary, the theoretical analysis confirms that the NSPDI module, through a sparse connection strategy, effectively controls both computational complexity and parameter count while maintaining excellent theoretical efficiency.

\section{Experiments}
\subsection{Datasets and Implementations}
We conducted comprehensive evaluations on three widely used benchmark datasets, including a gesture recognition event stream dataset (DVS128 Gesture~\citep{amir2017low}), a converted event-based dataset (CIFAR10-DVS~\citep{li2017cifar10}), and a static image classification dataset (CIFAR10~\citep{krizhevsky2009learning}). Additionally, we introduced speech recognition task (Spiking Heidelberg Digits, SHD~\citep{cramer2020heidelberg}) and reinforcement learning-based maze navigation task (Maze2d~\citep{github_project}) to further assess the model’s generalization in reinforcement learning contexts. 

For each event-based dataset, the input stream was segmented into a sequence of temporal slices of fixed length $\Delta t$, and only the first $T \times \Delta t$ events were retained per sample. For static image datasets, images were directly fed into the SNNs, where the first layer functioned as a temporal encoder, transforming static input into spiking activity.

All experiments were implemented using the \textbf{PyTorch} framework and executed on servers equipped with NVIDIA A100 GPUs. Unless otherwise noted, we adopted the default hyperparameter configuration outlined in Table~\ref{Tab1}, including the number of trials, epoch, batch size (BS), learning rate ($\eta$), learning rate decay schedule ($\gamma_{\eta}$), membrane time constant ($\tau_{\mathbf{m}}$), neuron firing threshold ($u_{\mathrm{th}}$), simulation time window ($T$), and integration timestep ($\Delta t$). In addition, we report two training-strategy parameters: $\text{Epoch}_{\text{C}}$ and $T_{\text{cyc}}$. Here, $\text{Epoch}_{\text{C}}$ denotes the additional epochs used after the main training phase to ensure final convergence (during these epochs $d_1$ and $d_2$ are fixed to their target values), and $T_{\text{cyc}}$ denotes the length (in epochs) of one cycle in the cyclic pruning schedule for $d_2$ (default $T_{\text{cyc}}=50$).

The pruning schedule for $d_1$ follows the sinusoidal scheme introduced in STDS. In practice, $d_1$ is expressed as a percentage. This directly determines the proportion of connections to prune, allowing precise and dynamic control of sparsity during training. The threshold $d_2$ is governed by a multi-periodic sine schedule inspired by cyclical pruning strategies~\citep{srinivas2022cyclical}. In our implementation, $d_2$ varies smoothly with epoch $e$ according to a (multi-)sinusoidal function with base period $T_{\text{cyc}}$ (default 50 epochs). This causes $d_2$ to progressively increase and decrease within each cycle. Additionally, at the end of the whole training run, we perform $\text{Epoch}_{\text{C}}$ extra epochs with $d_1$ and $d_2$ held at their final target values to guarantee convergence. This design makes pruning gradual and reversible within cycles while ensuring stability before the next cycle or at final convergence.

We employed task-specific SNN architectures across datasets, as detailed in Table~\ref{Tab2}. For instance, the network trained on DVS128 Gesture consisted of seven convolutional layers with the following structure: [Inputs-128C5S2-AP2-128C3-AP2-128C3-AP2-128C3-AP2-128C3-MP2-FC512-10], where “128C5S2” denotes a convolutional layer with 128 channels, $5 \times 5$ kernel, and stride of 2; “AP2” denotes $2 \times 2$ average pooling; and “FC512” indicates a fully connected layer with 512 output units. For the SHD dataset, dropout (DP) and batch normalization (BN) layers were incorporated to improve regularization.

Importantly, as the classification head plays a critical role in determining final performance, we excluded it from NSPDI pruning to preserve its full representational capacity. All reported performance metrics are the mean and standard deviation of top-1 accuracy over multiple independent runs, together with the corresponding best-performing scores.

\begin{figure*}[t]
	\centering
	\includegraphics{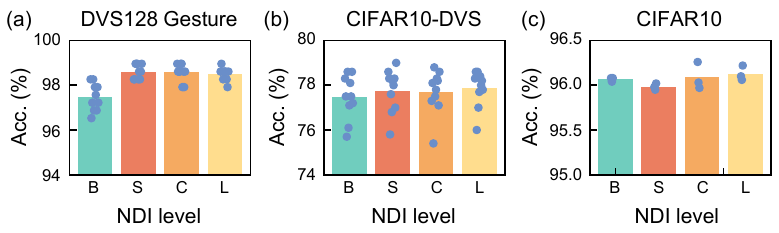}
	\caption{Average top-1 performance of different SNN models. Comparison of the baseline SNN (B, green) and three NDI-SNN variants—synapse-level (S, orange), channel-level (C, light orange), and layer-level (L, yellow)—on three datasets: (a) DVS128 Gesture, (b) CIFAR10-DVS, and (c) CIFAR10. Blue dots indicate the best top-1 accuracy achieved across independent trials.}
	\label{Fig3}
\end{figure*}

\begin{table*}[t]
	\renewcommand{\arraystretch}{1.}
	\caption{Average top-1 accuracies (upper) and best top-1 accuracies (bottom) achieved by the baseline SNN and the three-level NDI-SNN models on different datasets.}
	\label{Tab4}
	\centering
	\setlength{\tabcolsep}{5mm}{
		\begin{tabular}{c|cccc}
			\hline
			\multirow{2}*{Dataset}   & \multirow{2}*{Baseline-SNN} &                    &      NDI-SNN       &                    \\
			&                             &      Synapse-level       &      Channel-level       &       Layer-level        \\ \hline
			\multirow{1}*{DVS128} &     $97.47 \pm 0.58\%$      & $98.61 \pm 0.31\%$ & $98.58 \pm 0.36\%$ & $98.51 \pm 0.27\%$ \\
			Gesture          &          $98.26\%$          &     $98.96\%$      &     $98.96\%$      &     $98.96\%$      \\ \hline
			\multirow{1}*{CIFAR10} &     $77.47 \pm 0.94\%$      & $77.76 \pm 0.92\%$ & $77.71 \pm 0.93\%$ & $77.86 \pm 0.77\%$ \\
			-DVS             &          $78.60\%$          &     $79.90\%$      &     $78.80\%$      &     $78.60\%$      \\ \hline
			\multirow{2}*{CIFAR10}   &     $96.07 \pm 0.02\%$      & $95.98 \pm 0.03\%$ & $96.09 \pm 0.12\%$ & $96.13 \pm 0.07\%$ \\
			&          $96.08\%$          &     $96.02\%$      &     $96.26\%$      &     $96.22\%$      \\ \hline
		\end{tabular}
	}
\end{table*}

\begin{figure*}[t]
	\centering
	\includegraphics{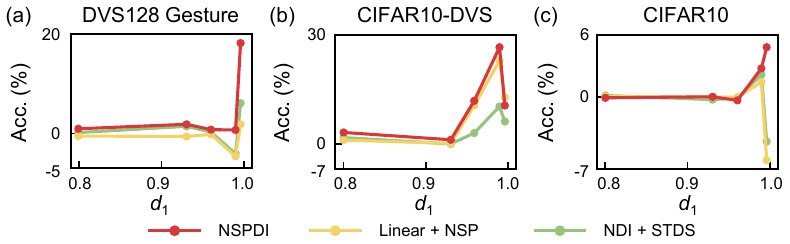}
	\caption{Accuracy difference with respect to the baseline sparse model (Linear + STDS) under varying sparsity levels on three datasets. Each line corresponds to one model: NSPDI (red), Linear + NSP(yellow), and NDI + STDS (green). (a) DVS128 Gesture dataset. (b) CIFAR10-DVS dataset. (c) CIFAR10 dataset.}
	\label{Fig4}
\end{figure*}

\subsection{NDI Improved Spatiotemporal Representation in SNNs}
NDI constitutes the core component of the proposed approach, designed to enhance the representational capacity and performance of SNNs. To evaluate the impact of NDI, we conducted controlled experiments comparing dense SNNs augmented with NDI against baseline SNNs without NDI across three benchmark datasets: DVS128 Gesture, CIFAR10-DVS, and CIFAR10.

Furthermore, we examined the influence of NDI at three structural levels: synapse-level, channel-level, and layer-level. Theoretically, NDI is designed to capture and integrate the spatiotemporal dependencies among inputs, which is especially advantageous for event-based datasets. However, because its coefficients are inherently linked to temporal dynamics, the benefits of NDI may diminish when applied to static image datasets that lack temporal structure.

As illustrated in Figure~\ref{Fig3}, including NDI significantly enhances model performance compared to the baseline across all evaluated datasets, with the most pronounced improvements observed on the event-driven DVS128 Gesture dataset. Quantitative results in Table~\ref{Tab4} show that at the synapse level, NDI improves the average top-1 accuracy from 97.47\% to 98.61\%, with the peak accuracy increasing from 98.26\% to 98.96\%. Similar performance gains are observed on CIFAR10-DVS, where synapse-level NDI boosts the maximum accuracy from 78.60\% to 79.90\%. These results highlight NDI's capability to model fine-grained spatiotemporal interactions, which is a critical factor for effectively processing event-based data.

In contrast, for the static CIFAR10 dataset, where temporal information is absent, the performance gains are comparatively modest. The average accuracy remains around 96\%, indicating that NDI's advantage is less pronounced in temporally invariant settings. This is because the effectiveness of NDI relies on the presence of temporal dynamics, where dendritic integration can capture and exploit cross-time dependencies. In purely static datasets such as CIFAR10, these spatiotemporal dependencies degenerate into spatial correlations only, thereby limiting the potential benefits of NDI. These findings collectively confirm the utility of NDI, particularly in domains rich in spatiotemporal complexity.

It is noteworthy that although synapse-level NDI achieved the highest average accuracy on DVS128 Gesture (98.61\%) and CIFAR10-DVS (78.60\%), its fine granularity results in an increased number of parameters. We ultimately selected channel-level NDI as the preferred configuration to maintain a balance between model performance and computational efficiency. This variant achieves comparable or even superior accuracy to synapse-level NDI across multiple datasets; for instance, it attains the highest accuracy of 96.26\% on CIFAR10 while incurring significantly fewer parameters. Thus, channel-level NDI offers a more practical and efficient solution for real-world deployment scenarios.

\begin{figure*}[!t]
	\centering
	\includegraphics{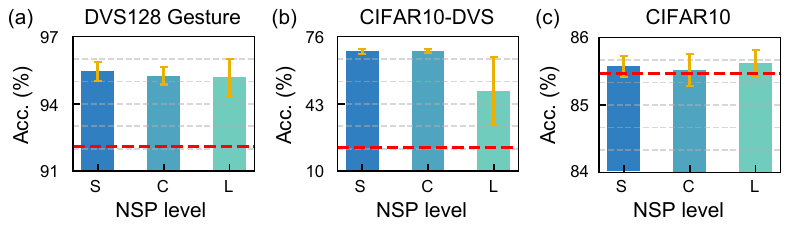}
	\caption{Accuracy comparison of NSPDI-SNN models using transition gains introduced at different structural levels: synapse-level (S, blue), channel-level(C, teal), and layer-level (L, green) levels. The red dashed line represents the accuracy of NDI-SNN using the STDS method. Each plot corresponds to one dataset: (a) DVS128 Gesture. (b) CIFAR10-DVS. (c) CIFAR10.}
	\label{Fig5}
\end{figure*}

\subsection{Ablation Study with NSP}
To enhance computational efficiency and reduce parameter redundancy in the high-performance NDI-SNN architecture, we introduce a lightweight pruning strategy termed NSP. NSP is an optimized extension of the STDS pruning framework, tailored to improve sparsity-aware training in spiking neural networks. To systematically evaluate the contribution of NSP under sparse learning, we applied both STDS and NSP to prune two model variants: the baseline SNN and the NDI-SNN. This yielded four distinct configurations: Linear + STDS, NDI + STDS, Linear + NSP, and NSPDI. The NDI method uses the channel level, and all pruning experiments were conducted with a fixed pruning schedule, where the initial pruning threshold $d_1$ was set to 100\%, and the enhancement threshold $d_2$ was set to 0.

Figure~\ref{Fig4} illustrates the accuracy improvements of Linear + NSP, NDI + STDS, and NSPDI, reported relative to the baseline sparse model (Linear + STDS) across three datasets: DVS128 Gesture, CIFAR10-DVS, and CIFAR10. The horizontal axis represents the retention weight ratio (0.8, 0.93, 0.96, 0.99, and 0.996), and the vertical axis represents the accuracy improvement relative to the baseline sparse model (Linear + STDS). NSP consistently outperforms STDS when applied to the baseline model, confirming its efficacy in enhancing sparse model performance. More notably, NSPDI achieves the highest accuracy across all evaluated sparsity levels. On the DVS128 Gesture dataset, NSPDI demonstrates a significant performance margin, particularly at higher sparsity levels, indicating its robustness in preserving spatiotemporal dynamics under structural compression. Similarly, on CIFAR10-DVS and CIFAR10, NSPDI maintains superior accuracy across nearly all sparsity regimes, underscoring the effectiveness of NSP in complementing the representational benefits introduced by NDI. These results validate the synergistic integration of NSP and NDI, highlighting NSPDI as a compact yet high-performing solution capable of balancing accuracy and computational efficiency in static and event-driven domains.

\subsection{Introducing the Transition Gain at Various Structural Levels}
The transition gain ($\bm{A}$) is the pivotal parameter governing heterogeneity in the proposed NSPDI-SNN framework. Adjusting the value of $\bm{A}$ modulates the magnitude of non-zero weights retained after reparameterization, thereby mitigating the loss of critical information induced by aggressive sparsification. To assess the effectiveness of transition gain integration across different architectural levels, we conducted experiments on three datasets using models with transition gain introduced at the synapse level ($\bm{A}$), channel (or neuron) level ($\bm{a}$), and layer level ($a$). In all experiments, we fixed the pruning thresholds as $d_1 = 99\%$ and $d_2 = 0$, ensuring consistent sparsity control for fair comparison.

Theoretically, synapse-level transition gain ($\bm{A}$) provides fine-grained control over sparsification, enabling per-neuron regulation. However, this granularity comes at the cost of a substantial increase in the number of learnable parameters. In contrast, channel-level ($\bm{a}$) and layer-level ($a$) transition gains offer coarser control, promoting global sparsity patterns with fewer additional parameters and lower computational overhead.

As illustrated in Figure~\ref{Fig5}, all three configurations demonstrate the capacity to alleviate accuracy degradation at increasing sparsity levels, highlighting the role of transition gain in enhancing sparse training dynamics. Furthermore, all achieve improved accuracy compared to the method using STDS (red dashed line). Among them, synapse-level transition gain yields the highest accuracy improvements but incurs significant computational costs due to excessive parameter growth and fine-grained regulation. Although lightweight in number of parameters, layer-level transition gain exhibits inconsistent performance, particularly under moderate-to-high sparsity levels, suggesting limited regulation capacity. In contrast, the channel-level transition gain ($\bm{a}$) consistently achieves a favorable trade-off between accuracy and complexity. It provides stable performance enhancements across all evaluated datasets and demonstrates superior robustness in the medium and high sparsity regimes. Consequently, we adopt the channel-level transition gain configuration in the final NSPDI-SNN design to ensure an optimal balance between pruning effectiveness and computational efficiency.

\begin{figure}[t]
	\centering
	\includegraphics{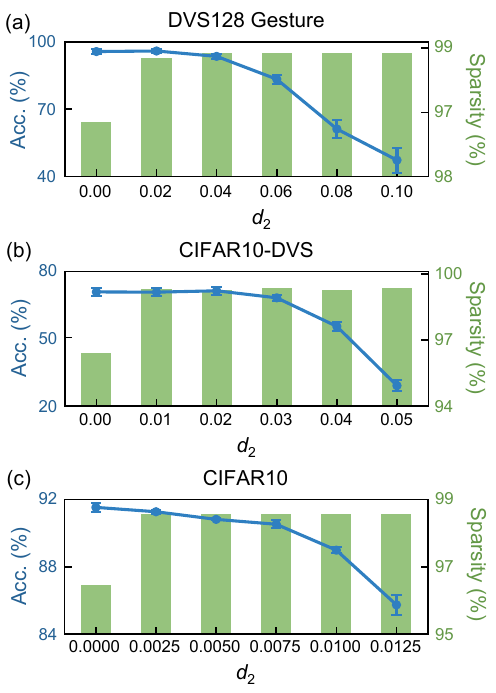}
	\caption{Accuracy (blue) and sparsity (green) exhibited by the NSPDI-SNN models under different values of $d_2$. (a) DVS128 Gesture. (b) CIFAR10-DVS. (c) CIFAR10.}
	\label{Fig6}
\end{figure}

\begin{figure}[t]
	\centering
	\includegraphics{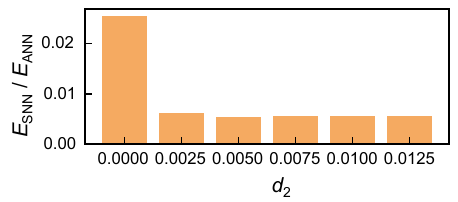}
	\caption{Energy ratio $E_{\text{SNN}}/E_{\text{ANN}}$ of NSPDI across different $d_2$ values on CIFAR-10.}
	\label{Fig7}
\end{figure}

\begin{figure*}[t]
	\centering
	\includegraphics{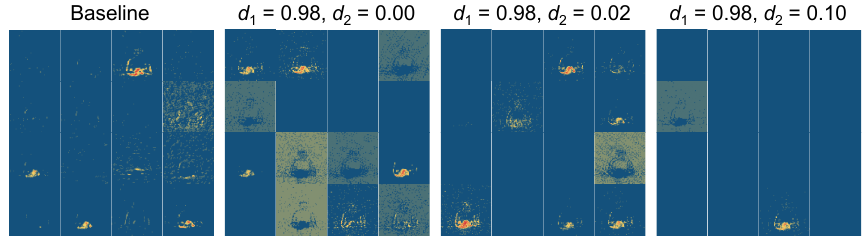}
	\caption{Visualization of the average spiking activity of the first convolutional layer on the DVS128 Gesture dataset across different models. The figure shows the feature maps of the first 16 output channels, where lighter colors indicate higher firing rates.}
	\label{Fig8}
\end{figure*}

\begin{table*}[!t]
	\renewcommand{\arraystretch}{1.}
	\caption{A detailed pruning comparison among the NSPDI-SNN model and other existing models on DVS128 Gesture, CIFAR10-DVS, and CIFAR10.} 
	\label{Tab5}
	\centering
	\begin{threeparttable}
		\setlength{\tabcolsep}{1mm}{
			\begin{tabular}{cccccccccc}
				\hline
				            \multirow{2}*{Dataset}             &                    Pruning                     &    \multirow{2}*{Arch.}     &               \multicolumn{2}{c}{Top-1 Acc. (Dense) (\%)}                &    Sparsity    &    \multirow{2}*{Params.}    & \multicolumn{2}{c}{Acc. Loss (\%)} & \multirow{2}*{T}  \\
				                                               &                     method                     &                             &                 Average                  &             Best              &      (\%)      &                              &    Average     &       Best        &                   \\ \hline
				\multirow{11}*{\rotatebox{90}{DVS128 Gesture}} &     \multirow{2}*{SCCD~\citep{Meng2022An}}     & \multirow{2}*{8 Conv, 1 FC} &             \multirow{2}*{-}             &     \multirow{2}*{94.44}      &     77.36      &     $40.00 \times 10^3$      &       -        &      -17.01       & \multirow{2}*{16} \\
				                                               &                                                &                             &                                          &                               &     90.13      &     $20.00 \times 10^3$      &       -        &      -19.09       &                   \\ \cline{2-10}
				                                               & \multirow{3}*{DC-STE~\citep{chen2023resource}} & \multirow{3}*{2 Conv, 2 FC} &             \multirow{3}*{-}             &     \multirow{3}*{93.75}      &     90.00      &      $1.25 \times 10^6$      &       -        &       -0.69       & \multirow{3}*{20} \\
				                                               &                                                &                             &                                          &                               &     93.00      &      $0.88 \times 10^6$      &       -        &       -0.69       &                   \\
				                                               &                                                &                             &                                          &                               &     95.00      &      $0.36 \times 10^6$      &       -        &       -1.04       &                   \\ \cline{2-10}
				                                               &   \multirow{3}*{STDS~\citep{chen2022state}}    & \multirow{3}*{2 Conv, 2 FC} &             \multirow{3}*{-}             &     \multirow{3}*{93.75}      &     90.00      &      $1.25 \times 10^6$      &       -        &       -4.17       & \multirow{3}*{20} \\
				                                               &                                                &                             &                                          &                               &     93.00      &      $0.88 \times 10^6$      &       -        &       -2.43       &                   \\
				                                               &                                                &                             &                                          &                               &     95.00      &      $0.36 \times 10^6$      &       -        &       -5.21       &                   \\ \cline{2-10}
				                                               &        \multirow{3}*{\textbf{Our work}}        & \multirow{3}*{5 Conv, 1 FC} & \multirow{3}*{\textbf{97.47 $\pm$ 0.58}} & \multirow{3}*{\textbf{98.26}} & \textbf{96.70} & \textbf{$20.12 \times 10^3$} & \textbf{-1.78} &  \textbf{-0.34}   & \multirow{3}*{8}  \\
				                                               &                                                &                             &                                          &                               & \textbf{98.70} & \textbf{$7.92 \times 10^3$}  & \textbf{-1.53} &  \textbf{-1.04}   &                   \\
				                                               &                                                &                             &                                          &                               & \textbf{98.86} & \textbf{$6.97 \times 10^3$}  & \textbf{-3.93} &  \textbf{-2.77}   &                   \\ \hline
				 \multirow{13}*{\rotatebox{90}{CIFAR10-DVS}}   &     \multirow{2}*{SCCD~\citep{Meng2022An}}     & \multirow{2}*{8 Conv, 1 FC} &             \multirow{2}*{-}             &     \multirow{2}*{72.60}      &     52.83      &      $0.56 \times 10^6$      &       -        &       -3.58       & \multirow{2}*{16} \\
				                                               &                                                &                             &                                          &                               &     80.82      &      $0.23 \times 10^6$      &       -        &       -5.96       &                   \\ \cline{2-10}
				                                               &   \multirow{3}*{STDS~\citep{chen2022state}}    &    \multirow{3}*{VGGSNN}    &             \multirow{3}*{-}             &      \multirow{3}*{82.4}      &     78.64      &      $3.52 \times 10^6$      &       -        &       -0.70       & \multirow{3}*{10} \\
				                                               &                                                &                             &                                          &                               &     89.83      &      $0.65 \times 10^6$      &       -        &       -1.30       &                   \\
				                                               &                                                &                             &                                          &                               &     95.33      &      $0.24 \times 10^6$      &       -        &       -2.60       &                   \\ \cline{2-10}
				                                               &  \multirow{3}*{UW-UN~\citep{shi2024towards}}   &    \multirow{3}*{VGGSNN}    &             \multirow{3}*{-}             &      \multirow{3}*{82.4}      &     95.54      &      $2.50 \times 10^6$      &       -        &       -1.40       & \multirow{3}*{10} \\
				                                               &                                                &                             &                                          &                               &     98.73      &      $2.18 \times 10^6$      &       -        &       -3.40       &                   \\
				                                               &                                                &                             &                                          &                               &     99.23      &      $1.81 \times 10^6$      &       -        &       -4.10       &                   \\ \cline{2-10}
				                                               &    \multirow{2}*{SCA~\citep{li2024towards}}     & \multirow{2}*{5 Conv, 1 FC} &             \multirow{2}*{-}             &     \multirow{2}*{72.80}      &     78.27      &      $0.25 \times 10^6$      &       -        &       +0.90       & \multirow{2}*{20} \\
				                                               &                                                &                             &                                          &                               &     93.05      &      $0.08 \times 10^6$      &       -        &       -0.90       &                   \\ \cline{2-10}
				                                               &        \multirow{3}*{\textbf{Our work}}        & \multirow{3}*{5 Conv, 1 FC} & \multirow{3}*{\textbf{77.47 $\pm$ 0.94}} & \multirow{3}*{\textbf{78.60}} & \textbf{94.04} & \textbf{$36.31 \times 10^3$} & \textbf{-4.25} &  \textbf{-4.10}   & \multirow{3}*{20} \\
				                                               &                                                &                             &                                          &                               & \textbf{98.84} & \textbf{$7.05 \times 10^3$}  & \textbf{-3.88} &  \textbf{-2.10}   &                   \\
				                                               &                                                &                             &                                          &                               & \textbf{98.94} & \textbf{$6.41 \times 10^3$}  & \textbf{-6.22} &  \textbf{-5.50}   &                   \\ \hline
				    \multirow{8}*{\rotatebox{90}{CIFAR10}}     &     \multirow{2}*{SCCD~\citep{Meng2022An}}     &  \multirow{2}*{ResNet-20}   &             \multirow{2}*{-}             &     \multirow{2}*{92.14}      &     37.94      &      $1.64 \times 10^6$      &       -        &       +0.22       & \multirow{2}*{4}  \\
				                                               &                                                &                             &                                          &                               &     80.36      &      $3.48 \times 10^6$      &       -        &       +1.11       &                   \\ \cline{2-10}
				                                               &   \multirow{3}*{ESL-SNN~\citep{shen2023esl}}    &  \multirow{3}*{ResNet-19}   &             \multirow{3}*{-}             &     \multirow{3}*{92.71}      &     67.29      &      $2.53 \times 10^6$      &       -        &       -0.63       & \multirow{3}*{2}  \\
				                                               &                                                &                             &                                          &                               &     83.94      &      $1.26 \times 10^6$      &       -        &       -1.25       &                   \\
				                                               &                                                &                             &                                          &                               &     91.79      &      $0.63 \times 10^6$      &       -        &       -1.81       &                   \\ \cline{2-10}
				                                               &        \multirow{3}*{\textbf{Our work}}        &   \multirow{3}*{ResNet19}   & \multirow{3}*{\textbf{96.07 $\pm$ 0.02}} & \multirow{3}*{\textbf{96.08}} & \textbf{96.45} & \textbf{$0.45 \times 10^6$}  & \textbf{-4.53} &  \textbf{-4.26}   & \multirow{3}*{4}  \\
				                                               &                                                &                             &                                          &                               & \textbf{98.57} & \textbf{$0.18 \times 10^6$}  & \textbf{-4.70} &  \textbf{-4.62}   &                   \\
				                                               &                                                &                             &                                          &                               & \textbf{98.58} & \textbf{$0.18 \times 10^6$}  & \textbf{-5.24} &  \textbf{-5.17}   &                   \\ \hline
			\end{tabular}}
	\end{threeparttable}
\end{table*}

\subsection{Sparsity Induced by the Pruning Threshold $d_2$}
To further enhance model efficiency and compress redundant parameters, we incorporated a threshold-based pruning mechanism within the NSPDI framework. This approach targets the elimination of low-magnitude synaptic weights, thereby increasing structural sparsity without significant loss of discriminative capacity. To systematically assess the influence of the pruning threshold $d_2$ on sparsity and performance, we conducted controlled experiments across three datasets: DVS128 Gesture, CIFAR10-DVS, and CIFAR10.

In all experiments, the NSPDI method was applied to sparsify the entire spiking neural network except for the final classification layer. We varied the threshold $d_2$ while keeping $d_1$ fixed to observe its effect on the final sparsity-accuracy trade-off. A cyclic pruning strategy was employed, where $d_2$ was progressively increased throughout training cycles; in the final stage of each cycle, $d_1$ and $d_2$ remained fixed until convergence. Specifically, we set $d_1 = 0.98$ for DVS128 Gesture and CIFAR10, and $d_1 = 0.96$ for CIFAR10-DVS, reflecting the different tolerance levels of each dataset to sparsification.

Figure~\ref{Fig6} presents the relationship between the pruning threshold $d_2$, final network sparsity, and classification accuracy. As $d_2$ increases, network sparsity rises accordingly, with minimal accuracy degradation observed at moderate sparsity levels (up to approximately 98\%). In this regime, the networks maintain stable performance, with only marginal reductions in accuracy relative to their dense counterparts. However, when $d_2$ surpasses critical thresholds (specifically, $d_2 > 0.08$ for DVS128 Gesture, $d_2 > 0.03$ for CIFAR10-DVS, and $d_2 > 0.01$ for CIFAR10), the models experience rapid performance deterioration. This trend is consistent with prior findings in sparse SNN research, where excessive pruning compromises the integrity of information flow by eliminating essential synaptic pathways. At extreme sparsity, only a limited number of active synapses remain, constraining feature propagation and degrading recognition capabilities.

To gain qualitative insights into the effect of $d_2$, we further analyzed the spiking activity within the convolutional layers. As shown in Figure~\ref{Fig7}, we visualized the first 16 channels of feature maps generated by the first convolutional layer in the SNN trained on DVS128 Gesture. Compared to the dense baseline, NSPDI-trained models exhibit significantly reduced spiking activity as $d_2$ increases. Notably, both the intensity and number of active feature maps decline, suggesting that the pruning process selectively retains the most informative synaptic connections. Given that DVS128 Gesture focuses primarily on recognizing dynamic hand gestures, these observations indicate that NSPDI effectively preserves task-relevant spatiotemporal features while discarding redundant connections.

In addition, we analyzed the energy efficiency of the NSPDI-SNN under different values of $d_{2}$ on the CIFAR10 dataset. We compared the energy ratio between the SNN and its ANN counterpart ($E_{\text{SNN}} / E_{\text{ANN}}$) at the same sparsity level and network structure. The theoretical energy cost for a multiply–accumulate (MAC) operation is estimated at 4.6~pJ, while that for an accumulate (AC) operation is 0.9~pJ~\citep{li2025brain}. As shown in Fig.~\ref{Fig7}, when $d_{2}$ increases from 0 to 0.0125, the energy ratio $E_{\text{SNN}} / E_{\text{ANN}}$ decreases from 0.026 to 0.005. Notably, the most significant drop occurs when $d_{2}$ increases from 0 to 0.0025, where the ratio sharply falls to 0.006. As  $d_{2}$ further increases, the additional computation introduced by NSPDI becomes dominant. In this case, the energy ratio is slightly affected by further pruning and a stable ratio appears for $d_{2} \geq 0.005$.

Collectively, these results highlight the critical role of the pruning threshold $d_2$ in controlling sparsity and maintaining functional capacity within the NSPDI-SNN architecture. Properly tuning $d_2$ enables the model to achieve high sparsity with minimal loss in performance, making it suitable for deployment in resource-constrained environments.

\subsection{Comparison with the State-of-the-Art Methods}
As shown in Table~\ref{Tab5}, we systematically evaluated the pruning performance of the proposed NSPDI-SNN model across multiple benchmark datasets. We compared it against several representative state-of-the-art~(SOTA) methods. The evaluation focuses on key aspects, including accuracy preservation, compression ratio (sparsity), number of parameters (Params.), and time window.

On the DVS128 Gesture dataset, NSPDI-SNN exhibits substantial advantages. While maintaining a high average Top-1 accuracy of 97.47\% and a peak accuracy of 98.26\%, the model demonstrates strong robustness under extreme pruning conditions. Even at a sparsity level of 98.86\%, the accuracy degradation is limited to only -2.77\%, which significantly outperforms prior methods such as SCCD~\citep{Meng2022An} and DC-STE~\citep{chen2023resource}. For example, SCCD suffers a -19.09\% performance drop at 90.13\% sparsity. The model achieves exceptional compactness and computational efficiency with fewer than 7K parameters (versus DC-STE). In the more challenging CIFAR10-DVS dataset, NSPDI-SNN continues to deliver competitive performance. Although the initial accuracy (78.60\%) is slightly lower than that of VGGSNN (82.4\%), NSPDI-SNN achieves superior robustness at extreme sparsity (98.94\%), with an accuracy loss of -5.5\%. Notably, NSPDI matches the accuracy of UW-UN~\citep{shi2024towards} but with just 7.05K parameters (versus multi-million in UW-UN), drastically cutting computational costs and enabling real-world deployment. For the static image dataset CIFAR10, NSPDI-SNN is pruned based on the spiking ResNet19 backbone. The model achieves an initial accuracy of 96.08\%, with a maximum sparsity of 98.58\% and a worst-case accuracy drop of only -5.17\%. This performance exceeds that of ESL-SNN~\citep{shen2023esl}, which incurs a -1.81\% accuracy loss at 91.79\% accuracy while still retaining 0.63M parameters. In contrast, NSPDI compresses the parameter count to under 0.18M, significantly reducing model size and resource demands.

In summary, the proposed NSPDI-SNN achieves state-of-the-art performance on event-driven datasets while maintaining effective classification accuracy under high sparsity in static image tasks. The synergistic integration of NSPDI demonstrates both strong practical value and broad adaptability in scenarios requiring efficient and compact neuromorphic computation.
\begin{table}[!t]
	
	\renewcommand{\arraystretch}{1.}
	\caption{Performance of NSPDI-SNN on the SHD dataset.} 
	\label{Tab6}
	\centering
	\begin{threeparttable} 
		\setlength{\tabcolsep}{1.2mm}{
			\begin{tabular}{cccccc}
				\hline
				Pruning         &             \multicolumn{2}{c}{Acc. (\%) }             & Sparsity & \multicolumn{2}{c}{Acc. Loss (\%) } \\
				method          &             Average             &         Best         &   (\%)   & Average &           Best            \\ \hline
				STDS\tnote{1}      &        $74.21 \pm 1.05$         &        76.33         &  93.72   &  -3.82  &           -0.23           \\ \hline
				\multirow{5}*{Our work} & \multirow{5}*{$74.21 \pm 1.05$} & \multirow{5}*{76.33} &  93.47   &  +8.26  &           +8.61           \\
				&                                 &                      &  93.86   &  +8.71  &           +8.74           \\
				&                                 &                      &  94.12   &  +8.17  &           +8.12           \\
				&                                 &                      &  94.76   &  +7.06  &           +6.80           \\
				&                                 &                      &  95.21   &  +3.73  &           +5.25           \\ \hline
		\end{tabular}}
		\begin{tablenotes}
			\footnotesize
			\item[1] Our implementation.
		\end{tablenotes}
	\end{threeparttable}
\end{table}

\begin{table}[!t]
	\renewcommand{\arraystretch}{1.25}
	\caption{Influence of sparsity on maze navigation performance.} 
	\label{Tab7}
	\centering
	\begin{threeparttable}
		\setlength{\tabcolsep}{4mm}{
			\begin{tabular}{cccc}
				\hline
				          Pruning            & Sparsity & \multirow{2}*{Avg. step} & \multirow{2}*{Reward} \\
				           method            &   (\%)   &                          &                       \\ \hline
				\multirow{3}*{STDS\tnote{1}} &   0.00   &          160.40          &        4375.84        \\
				                             &  19.79   &          188.20          &       -5214.80        \\
				                             &  39.68   &          200.00          &       -4374.98        \\ \hline
				  \multirow{5}*{Our work}    &   0.00   &          83.30           &        6979.09        \\
				                             &  19.14   &          126.30          &        4056.43        \\
				                             &  38.93   &          138.40          &        1760.85        \\
				                             &  58.41   &          141.30          &        858.66         \\
				                             &  79.55   &          174.30          &        1051.94        \\ \hline
			\end{tabular}}
		\begin{tablenotes}
			\footnotesize
			\item[1] Our implementation.
		\end{tablenotes}
	\end{threeparttable}
\end{table}

\begin{figure*}[t]
	\centering
	\includegraphics{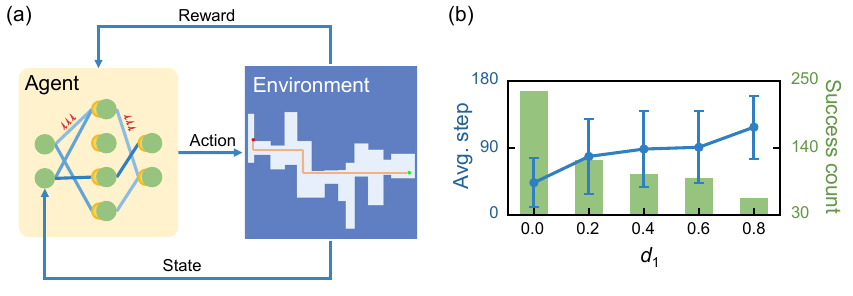}
	\caption{Application and effect evaluation of the NSPDI method in maze navigation tasks. (a) shows the overall structure of the deep Q-learning method built on the PyTorch-Maze-Solving framework after the introduction of the NSPDI mechanism. (b) describes the changing trend of the average number of successful steps (Avg. step) and the number of successes (Success count) of the agent in the maze environment under different pruning levels.}
	\label{Fig9}
\end{figure*}

\subsection{Validation on Complicated Task}
As we further explore the NSPDI framework, its flexibility and adaptability across diverse application scenarios become increasingly evident. Beyond standard classification tasks, NSPDI-SNN demonstrates strong generalization capabilities in more complex, real-world environments, such as speech recognition and maze navigation. In these scenarios, the model is required to make integrated decisions involving temporal dynamics, spatial positioning, and contextual information processing.

\subsubsection{Speech Recognition on the SHD Dataset}
To evaluate the effectiveness of NSPDI in speech recognition tasks, we conducted experiments on the SHD dataset. A simple spiking neural network comprising three fully connected layers was used as the backbone architecture. To concentrate on compressing redundant parameters in the feature extraction stages, the NSPDI mechanism was applied exclusively to the layers preceding the final classification layer. During training, we fixed the initial pruning threshold to $d_1 = 0.96$ and gradually increased the sparsity enhancement threshold $d_2$ to generate model variants with varying sparsity levels. This allowed for a systematic investigation of the impact of different sparse structures on recognition performance.

As shown in Table~\ref{Tab7}, NSPDI achieves competitive accuracy while significantly improving model sparsity. The best-performing model reached a sparsity of 95.68\% with only a +2.65\% change in accuracy relative to the baseline, confirming that NSPDI can maintain or even enhance performance despite aggressive parameter reduction. Compared with the STDS-based baseline, NSPDI demonstrates better parameter efficiency and competitive classification accuracy across different sparsity configurations, highlighting its utility in low-resource neuromorphic speech processing scenarios.

\subsubsection{Decision-Making in Maze Navigation Tasks}
To further validate the generalization ability of NSPDI, we extended our evaluation to a reinforcement learning setting using a discrete maze navigation task. The agent was trained using the Deep Q-Network (DQN) algorithm, where the Q-function was approximated by a three-layer fully connected neural network, as outlined in Table~\ref{Tab3}. The current position of the agent served as network input, while the output layer produced Q-values for four discrete actions (up, down, left, right). The reward scheme consisted of +5 for goal attainment, -0.3 for stationary actions, and $(1-d)$ otherwise, where $d$ denotes the normalized target distance. We employed experience replay (buffer size=100k, batch size=100) and $\epsilon$-greedy exploration ($\epsilon$ annealed from 0.9 to 0.01), with optimization via Adam ($\eta=10^{-3}$).

Figure~\ref{Fig9}(b) and Table~\ref{Tab5} summarize the results. Under non-pruned conditions (0.00\% sparsity), the NSPDI agent required an average of only 83.30 steps to reach the target, significantly fewer than the 160.40 steps required by STDS, indicating superior planning efficiency. Moreover, NSPDI achieved a reward of 6979.09, surpassing STDS’s 4375.84. As sparsity increased, NSPDI retained stable performance up to moderate sparsity levels. Even at 79.55\% sparsity, the average steps increased to only 174.30, which remains lower than STDS at 39.68\% sparsity (200.00 steps). The corresponding reward also remained higher than STDS, illustrating the resilience of NSPDI to performance degradation under extreme pruning. These findings, reinforced by the trends shown in Figure~\ref{Fig9}(b), indicate that NSPDI achieves a favorable balance between computational efficiency and behavioral performance. It demonstrates both pruning robustness and effective strategy learning in reinforcement learning tasks.

\section{Discussion and Conclusion}
In this work, we proposed NSPDI, a biologically inspired and computationally efficient lightweight learning framework for SNNs. The core of our approach lies in integrating two synergistic components: NDI and NSP. Specifically, NDI augments LIF neurons with nonlinear dendritic computations that encode spatiotemporal dependencies among spikes inspired by dendritic processes in biological neurons. Meanwhile, NSP reparameterizes synaptic weights using a learnable transition gain, extending the STDS framework and enabling finer-grained control over synaptic sparsity. Through the combination of NDI and NSP, our NSPDI framework facilitates high-performance sparse learning in SNNs, offering an effective and scalable solution suitable for neuromorphic and edge deployment scenarios.

We conducted extensive and systematic experiments on three benchmark datasets (DVS128 Gesture, CIFAR10-DVS, and CIFAR10) to evaluate the efficacy of NSPDI. Our ablation studies demonstrated that NDI significantly improves the representation of spatiotemporal patterns, thereby enhancing classification performance. We further validated the moderating role of NSP in sparsification by analyzing the effects of channel-level transition gain ($\bm{a}$) and pruning threshold ($d_2$), revealing that NSPDI achieves controllable sparsification with minimal performance degradation. Moreover, we extended our evaluation to two additional tasks: the SHD speech recognition dataset and a reinforcement learning-based maze navigation task. In both cases, NSPDI exhibited superior robustness and generalization under varying sparsity levels. These results confirm the broad applicability and effectiveness of our method in diverse SNN application domains.

The biological interpretability of NSPDI is rooted in its principled design inspired by neuronal mechanisms observed in cortical structures such as Purkinje cells. NDI simulates the nonlinear dendritic computations that enhance synaptic integration and feature selectivity~\citep{hao2009arithmetic, li2014bilinearity, li2019dendritic}. NSP, on the other hand, draws from the concept of dynamic synaptic organization by mimicking the state transition ratio between dendritic spines and filopodia~\citep{chen2022state}, a phenomenon linked to synaptic development and pruning. By combining dendritic nonlinearity with multilevel heterogeneity in synaptic transition~\citep{dobrunz1997heterogeneity, fritschy2012molecular, perez2021neural}, our framework embodies a biologically plausible model that translates structural and functional aspects of real neural circuits into effective sparse learning in artificial SNNs. Notably, our final design choice of channel-level NDI and pruning reflects a balance between biological fidelity, performance gains, and computational tractability.

Despite its promising results, several limitations remain. The introduction of NDI introduces additional computational overhead, particularly in the case of fine-grained (e.g., synapse-level) modeling. We alleviated this by adopting channel-level implementations, and note that the event-driven, sparse AC operations of SNNs still provide a fundamental efficiency advantage over the dense MAC operations of ANNs (especially in fully connected layers), but this advantage is partially offset in our current architecture by the use of average pooling in convolutional layers. Future work should explore more efficient approximations or hardware-friendly variants of nonlinear integration. Replacing average pooling with max pooling can be efficiently realized through logical OR operations on spike events. This modification presents a clear path to further amplify energy efficiency. Additionally, integration with complementary learning strategies such as spike-based attention, local learning rules, or meta-plasticity mechanisms may further enhance the robustness and generalization of sparse SNNs.

\section*{Acknowledgments}
This work was supported in part by the National Key Research and Development Program of China under grant 2023YFF1204200, in part by the STI 2030–Major Projects under grant 2022ZD0208500, in part by the Sichuan Science and Technology Program under grant 2024NSFTD0032, grant 2024NSFJQ0004 and grant DQ202410, and in part by the Natural Science Foundation of Chongqing, China under grant CSTB2024NSCQ-MSX0627, in part by the Science and Technology Research Program of Chongqing Education Commission of China under Grant KJZD-K202401603, and in part by the China Postdoctoral Science Foundation under grant 2024M763876.

\bibliographystyle{elsarticle-harv} 
\bibliography{ref.bib}

\end{document}